\documentclass{amsart}
\usepackage[utf8]{inputenc}
\usepackage{amsmath}
\usepackage{amsfonts}
\usepackage{amssymb}
\usepackage{graphicx}
\usepackage{booktabs}

\numberwithin{equation}{section}  
\newtheorem{defn}{Definition}[section]

\newtheorem{rem}[defn]{Remark}
\newtheorem{thm}[defn]{Theorem}

\newtheorem{lem}[defn]{Lemma}

\title
{Some Computations for Optimal Execution with Monotone Strategies}

\author{Yan Dolinsky}
\address{Department of Statistics, Hebrew University, Jerusalem, Israel}
          
\begin{document}
\vspace{-0.5cm}

\begin{abstract}
We study
an optimal execution problem in the infinite horizon setup. 
Our financial market is given by the Black-Scholes model with a linear price impact. 
The main novelty of the current note is that we study the constrained case where the 
number of shares and the selling rate are non-negative processes. 
For this case we give a complete characterization
of the value and the optimal control via a solution of a non-linear ordinary differential equation (ODE).
Furthermore, we provide an example where the non-linear ODE can be solved explicitly. 
Our approach is purely probabilistic. 
\end{abstract}


\begin{minipage}[t]{\textwidth}
\vspace*{-4\baselineskip}
\maketitle
\end{minipage}

\section{Introduction and the Main Result}\label{sec:2}
In financial markets, 
trading moves prices against the trader:
buying faster increases execution prices, and selling faster decreases them. This aspect
of liquidity, known as market depth \cite{B} or price impact 
received increasing attention in models of illiquid portfolio choice.
One of the most widely used models of illiquidity go back to Almgren and Chriss \cite{AC}.
Loosely speaking, the
model of Almgren and Chriss is characterized by directly specifying functions
describing the temporary and permanent impacts of a given order on the
price. 

In \cite{GS:11} the authors solved analytically 
the optimal execution problem for the
case where the risky asset is given by 
geometric Brownian motion
and trading is subject to a linear price impact. 
The authors did not pose any constraints on the trading strategy. 
In particular, in their setup, the number of shares and the selling rate can take on negative values. 
In real market conditions, there may be various constraints on the trading strategies. These constraints
help to prevent excessive trading that could destabilize the market or to comply with regulations. 

The current work treats the infinite horizon version 
of \cite{GS:11} under the additional constraint that the trading rate and the inventory process
are non-negative. The constrained setup is much more technical (than the setup in \cite{GS:11}) and so, in order to have a tractable solution,
 we consider the 
perpetual case which allows to reduce the dimension of the corresponding optimization problem.
It is important to mention the recent preprint \cite{ADH} which provides an algorithmic tool 
to solve optimal execution problems with various types of constraints. Still, as far as we know, the current paper is the first to provide 
an analytical solution to a constrained optimal execution problem.

Next, we formulate our results.
Let 
$W=\left(W_t\right)_{t \geq 0}$
be a standard Brownian motion defined on the filtered probability space
$(\Omega, \mathcal{F},(\mathcal F_t)_{t\geq 0}, \mathbb P)$
where the filtration $(\mathcal F_t)_{0\leq t\leq T}$ satisfies the
usual assumptions (right-continuity and completeness).
Consider a simple
financial market with a riskless savings account bearing zero interest (for simplicity)
and with a risky asset whose price is given by a geometric Brownian motion 
\begin{equation}\label{1}
S_t=S_0\exp\left(\sigma W_t+\left(\mu-\frac{\sigma^2}{2}\right)t\right), \ \ t\geq 0,
\end{equation}
where $S_0>0$ is the initial stock price, $\sigma>0$ is the constant volatility and $\mu<0$ is the constant negative drift. 
Negative drift means that $S$ is a super-martingale. 

Following \cite{AC}, we model the investor’s market impact in a
temporary linear form and thus, when at time $t$ the investor turns over her position $\Phi_t$ at the rate $\phi_t=\dot{\Phi}_t$
the execution price is $S_t+\frac{\Lambda}{2}\phi_t$ for some constant $\Lambda>0$. 
In our setup the number of shares $\Phi=(\Phi_t)_{t\geq 0}$ and the selling rate $-\phi=(-\phi_t)_{t\geq 0}$ are nonnegative processes.
As a result, the infinite horizon profits are given by 
\begin{equation}\label{2}
V^{\phi,S_0}:=-\int_{0}^{\infty} \phi_t S_t dt-\frac{\Lambda}{2}\int_{0}^{\infty} \phi^2_t dt.
\end{equation}
Let $\mathbb E$ denote the expectation with respect to $\mathbb P$. 
Assume that the initial number of shares $\Phi_0>0$ is a given deterministic number. Since $\phi\leq 0$ we obtain
\begin{align}
&\mathbb E\left[-\int_{0}^{\infty} \phi_t S_t dt\right]=\lim_{T\rightarrow\infty} \mathbb E\left[-\int_{0}^{T} \phi_t S_t dt\right]\nonumber\\
&\leq S_0 \lim_{T\rightarrow\infty} \mathbb E\left[-\int_{0}^{T} \phi_t \exp\left(\sigma W_t-\frac{\sigma^2}{2}t\right) dt\right]\nonumber\\
&=S_0\lim_{T\rightarrow\infty} \mathbb E\left[-\exp\left(\sigma W_T-\frac{\sigma^2}{2}T\right) \int_{0}^{T} \phi_t dt\right]\leq \Phi_0 S_0.\label{1+}
\end{align}
Indeed, the first equality follows from the monotone convergence theorem. The first inequality is due to $\mu\leq 0$. 
The second equality follows from the martingale property. The last inequality follows from 
$\Phi_0+\int_{0}^T \phi_t dt\geq 0$ for all $t$.

We conclude that 
$-\int_{0}^{\infty} \phi_t S_t dt$ is finite almost surely (a.s.) and so,
the right hand side of (\ref{2})
is well defined (may get the value $-\infty$). Thus, for $\Phi_0>0$ denote by 
$\mathcal A_{\Phi_0}$ the set of all 
$(\mathcal F_t)_{t\geq 0}$-progressively measurable processes $\phi=(\phi)_{t\geq 0}$ which satisfy 
$-\phi_t\leq 0$ $dt\otimes\mathbb P$ a.s and 
$\Phi_0+\int_{0}^{\infty}\phi_t dt\geq 0$ a.s.
We are interested in the following infinite horizon optimal execution problem 
\begin{equation}\label{3}
G(\Phi_0,S_0):=\sup_{\phi\in\mathcal A_{\Phi_0}}\mathbb E\left[V^{\phi,S_0}\right].
\end{equation}

Consider the bijection $\mathcal A_{\Phi_0}\rightarrow \mathcal A_{\frac{\Phi_0}{S_0}}$ given by
$\phi\rightarrow\frac{\phi}{S_0}$. From (\ref{1}) we get 
$V^{\phi,S_0}=S^2_0 V^{\frac{\phi}{S_0},1}.$ Hence, 
$G(\Phi_0,S_0)=S^2_0 G\left(\frac{\Phi_0}{S_0},1\right)$. We arrive at the main result which will be proved in Section 2.
\begin{thm}\label{thm2.1}
The value function is given by 
$G(\Phi_0,S_0)=S^2_0 g\left(\frac{\Phi_0}{S_0}\right)$ where $g:[0,\infty)\rightarrow [0,\infty)$ is the unique function which satisfies the following properties: \\
(i) The function $g$ is non-decreasing and concave with the initial conditions  $g(0)=0$ and $g'(0)=1$.\\
(ii) The function $g$ solves the non-linear ODE 
\begin{equation}\label{4}
\frac{\sigma^2 x^2 }{2}g''(x)-(\mu+\sigma^2) xg'(x)+(2\mu+\sigma^2) g(x)+\frac{\left(1-g'(x)\right)^2}{2\Lambda }=0, \ \ x>0.
\end{equation}

Moreover, the trading strategy $\hat\phi\in\mathcal A_{\Phi_0}$  which is given by the ODE 
\begin{equation}\label{5}
\hat\phi_t:=\frac{d\hat\Phi_t}{dt}=-S_t\left(1-g'\left(\frac{\hat\Phi_t}{S_t}\right)\right), \ \ t\geq 0
\end{equation}
is the unique ($dt\otimes\mathbb P$ a.s.) optimal trading strategy. 
\end{thm}
We end this section with several remarks.

\begin{rem}
Observe that the non-linear ODE given by (\ref{4}) has a singularity at $x=0$, and so 
given the initial conditions $g(0)=0$ and $g'(0)=1$,
the existence and the uniqueness of a solution can not 
be established by standard arguments. We will show by  completely probabilistic arguments that if we add the requirement 
that $g$ is non-decreasing and concave, then we get a unique solution. 
Let us notice that for any interval $(a,b)$, a solution to the ODE (\ref{4}) can not take a constant value in $(a,b)$.
Hence, in our case the non-decreasing property is equivalent to the solution being strictly increasing.
\end{rem}

\begin{rem}
From properties (i)-(ii) in Theorem \ref{thm2.1}
it follows that the function $g':(0,\infty)\rightarrow[0,1]$ is locally Lipschitz continuous, and so there exists 
a unique solution (see Chapter II in \cite{W}) to the ODE (\ref{5}) up to the hitting time $\inf\{t:\hat\Phi_t=0\}$. 
After this hitting time we put $\hat\phi\equiv 0$. 
Hence, the ODE given by (\ref{5}) is well defined and provides a unique trading strategy.
From the fact that $g'\geq 0$ we obtain that $-\phi\leq S$ and so 
$\Phi_t\geq \Phi_0-\int_{0}^{t}S_u du$. Since $\mu<0$, the integral
$\int_{0}^{\infty} S_u du<\infty$ a.s. and its distribution  
has a full support on $\mathbb R_{+}$ (see \cite{D}). Hence, $\mathbb P(\lim_{t\rightarrow\infty}\Phi_t>0)>0$,
and in particular the hitting time $\inf\{t:\hat\Phi_t=0\}$ can take the value $\infty$ with a positive probability. 
\end{rem}
\begin{rem}
For the case of a positive drift $\mu>0$, the value of the optimization problem given by (\ref{3}) is infinity. 
Indeed consider the trading strategy $\phi\in\mathcal A_{\Phi_0}$ given by 
$\phi_t=-\mu \Phi_0   e^{-\mu t}$, $t\geq 0$. Then
$\mathbb E[-\phi_t S_t]=\mu\Phi_0$ for all $t$ and 
$\int_{0}^{\infty} \phi^2_t dt=\frac{\mu\Phi^2_0}{2}$. 
Hence $\mathbb E\left[V^{\phi,S_0}\right]=\infty$.

If $\mu=0$, i.e. $S$ is a martingale. Then, (similarly to (\ref{1+}))
$$\mathbb E\left[-\int_{0}^{\infty} \phi_t S_t dt\right]\leq \Phi_0 S_0.$$
Thus, for any $\phi\in\mathcal A_{\Phi_0}$ we have 
 $\mathbb E\left[V^{\phi,S_0}\right]<\Phi_0 M_0$.
 On the other hand, for the sequence of trading strategies $\phi^n\in\mathcal A_{\Phi_0}$, $n\in\mathbb N$ given by 
 $\phi^n_t:=-\frac{\Phi_0e^{-t/n}}{n}$, $t\geq 0$ we get 
 $\lim_{n\rightarrow\infty}\mathbb E\left[V^{\phi^n,S_0}\right]=\Phi_0 S_0.$
 We conclude that for the case where $S$ is a martingale the value function is given by 
 $G(\Phi_0,S_0)=\Phi_0 S_0$ and there is no optimal control. 
\end{rem}

\subsection{\textbf{Example with an Explicit Solution}}
In this section we treat the case where $2\mu+\sigma^2=0$, i.e. $S^2$ is a martingale. 
Consider the change of variables  $y:=\frac{1}{x}$
and let $h(y):=1-g'(x)$ for $x>0$. Then (\ref{4}) is equivalent to 
the
Riccati equation
\begin{equation}\label{ric}
h'(y)+\frac{h(y)}{y}+\frac{h^2(y)}{\Lambda\sigma^2}=\frac{1}{y}.
\end{equation}
Substituting  $h(y):=\Lambda\sigma^2\frac{v'(y)}{v(y)}$ we obtain the ODE
\begin{equation}\label{ODE}  
y v''(y)+v'(y)-\frac{1}{\sigma^2\Lambda} v(y)=0.
\end{equation}
The general solution of this equation is given by 
$$v(y)=C_1 J_0\left(\frac{2i\sqrt y}{\sigma\sqrt\Lambda}\right)+C_2 Y_0\left(\frac{2i\sqrt y}{\sigma\sqrt\Lambda}\right)$$
where $i=\sqrt {-1}$, $C_1,C_2\in\mathbb R$ are arbitrary constants and 
$J_0$ and $Y_0$ are the Bessel functions (of order zero) of the first and the second kind. 
For details see 
Chapter II and Supplement II in \cite{PZ}.

From the definition 
$J_0\left(\frac{2i\sqrt y}{\sigma\sqrt\Lambda}\right):=\sum_{n=0}^{\infty}\frac{\left(\frac{y}
{\Lambda \sigma^2}\right)^n}{n!^2}.$ 
 Set,
 $$h(y):=\Lambda\sigma^2\frac{\frac{d}{dy}J_0\left(\frac{2i\sqrt y}
 {\sigma\sqrt\Lambda}\right)}{J_0\left(\frac{2i\sqrt y}{\sigma\sqrt\Lambda}\right)}=
 \frac{\sum_{n=0}^{\infty}\frac{\left(\frac{y}
{\Lambda \sigma^2}\right)^n}{(n+1)
n!^2}}{\sum_{n=0}^{\infty}\frac{\left(\frac{y}
{\Lambda \sigma^2}\right)^n}{n!^2}}
 , \ \ \ y>0.$$

Observe that 
$\lim_{y\rightarrow 0} h(y)=1$ and
$\lim_{y\rightarrow\infty} h(y)=0$. Let us argue that $h$ is non-increasing. Assume by contradiction that this is not the case, then 
from the above limits and the fact that $h\in (0,1)$ it follows that there exist $y_1<y_2$ such that $y_1$
is a local minimum  of $h$, $y_2$ is a local maximum of $h$ and $h(y_2)> h(y_1)$. This together with (\ref{ric}) gives 
$$\frac{h^2(y_1)}{\Lambda\sigma^2}=\frac{1-h(y_1)}{y_1}>\frac{1-h(y_2)}{y_2}=\frac{h^2(y_2)}{\Lambda\sigma^2}>
\frac{h^2(y_1)}{\Lambda\sigma^2}$$ and we get a contradiction. Thus $h$ is non-increasing.

We conclude that the function 
$g(x)=\int_{0}^x \left(1-h'\left(\frac{1}{z}\right)\right)dz$
satisfies the properties (i)-(ii) which are given in Theorem \ref{thm2.1}.
 Hence, the optimal control satisfies the ODE 
$$\hat\phi_t:=\frac{d\hat\Phi}{dt}=-S_t \frac{\sum_{n=0}^{\infty}\frac{\left(\frac{S_t}{\Lambda\sigma^2 \hat\Phi_t}\right)^n}{(n+1)n!^2}}
 {\sum_{n=0}^{\infty}\frac{\left(\frac{S_t}{\Lambda\sigma^2 \hat\Phi_t}\right)^n}{n!^2}}, \ \ \ t\geq 0$$
 and the value function is given by 
 $$G(\Phi_0,S_0)=S^2_0 g\left(\frac{\Phi_0}{S_0}\right)=\Phi_0 S_0-S^2_0\int_{0}^{\frac{\Phi_0}{S_0}}\frac{\sum_{n=0}^{\infty}\frac{\left(\frac{1}{\Lambda\sigma^2 x}\right)^n}{(n+1)n!^2}}
 {\sum_{n=0}^{\infty}\frac{\left(\frac{1}{\Lambda\sigma^2 x}\right)^n}{n!^2}}  dx.$$

\section{Proof of Theorem \ref{thm2.1}}\label{sec:3}
\begin{lem}\label{lem1}
There exists a unique optimal control to the optimization problem given by (\ref{3}). 
\end{lem}
\textbf{Proof:}
Clearly, the set $\mathcal A_{\Phi_0}$ is a convex set. Thus, uniqueness follows from the strict concavity of the functional
$\phi\rightarrow \mathbb E\left[V^{\phi,S_0}\right]$.

Next, we apply the Komlos lemma for proving existence. 
Let $\phi^n\in\mathcal A_{\Phi_0}$ $n\in\mathbb N$ be a sequence of 
trading strategies such that 
$G(\Phi_0,S_0)=\lim_{n\rightarrow\infty} \mathbb E\left[V^{\phi^n,S_0}\right].$
By lemma A1.1. in \cite{DS} there exists a sequence 
$\eta^n\in conv (\phi^n,\phi^{n+1},...)$ such that $\eta^n$ converge $dt\otimes\mathbb P$ almost surely to a limit 
$\eta\leq  0$. Let us show that $\eta$ is an optimal control.

From the concavity of the map $\phi\rightarrow \mathbb E\left[V^{\phi,S_0}\right]$ we get 
 \begin{equation}\label{7}
 G(\Phi_0,S_0)=\lim_{n\rightarrow\infty} \mathbb E\left[V^{\eta^n,S_0}\right].
 \end{equation}
As in (\ref{1+}) we have $\mathbb E\left[-\int_{0}^{\infty} \eta^n_t S_t dt\right]\leq \Phi_0 S_0$ for all $n\in\mathbb N$.
Hence,
\begin{equation}\label{7+}
\sup_{n\in\mathbb N}\mathbb E\left[\int_{0}^{\infty}|\eta^n_t|^2 dt\right]<\infty.
\end{equation}

The Fatou lemma implies 
\begin{equation}\label{8}
\mathbb E\left[\int_{0}^{\infty}\eta^2_t dt\right]\leq \lim\inf_{n\rightarrow\infty}\mathbb E\left[\int_{0}^{\infty}|\eta^n_t|^2 dt\right]
\end{equation}
and 
$\int_{0}^\infty -\eta_t dt\leq \lim\inf_{n\rightarrow\infty}-\eta^n_t dt\leq \Phi_0.$
In particular $\eta\in\mathcal A_{\Phi_0}$.

From (\ref{7+}) it follows that for any $T>0$ and $p\in (1,2)$  
$$\lim_{n\rightarrow\infty}\mathbb E_{\mathbb P}\left[\int_{0}^T |\eta^n_t-\eta_t|^p dt \right]=0.$$
Hence, from the Hölder's inequality 
\begin{equation}\label{9}
\lim_{n\rightarrow\infty}\mathbb E\left[\int_{0}^{T}|\eta^n_t-\eta_t| S_t dt\right]=0, \ \ \forall T>0.
\end{equation}

Next, by using the same arguments as in (\ref{1+}) it follows that for any trading strategy $\phi\in \mathcal A_{\Phi_0}$ 
and $T>0$ we have the uniform bound 
$$
\mathbb E\left[-\int_{T}^{\infty} \phi_t S_t dt\right]\leq \mathbb E\left[\Phi_T S_T\right]\leq \Phi_0 \mathbb E\left[S_T\right]=\Phi_0 e^{\mu T}.
$$
This together with 
(\ref{7}) and (\ref{8})-(\ref{9}) (we take $T\rightarrow\infty$) yields that $\eta$ is the optimal control. 
\qed
${}$\\
${}$\\
Next, we establish the properties of the function $g$.
\begin{lem}\label{lem2}
The function $g:[0,\infty)\rightarrow [0,\infty)$ is a non-decreasing, concave function which satisfies 
$g(0)=0$ and $g'(0)=1$. Moreover, for
for any $\epsilon>0$ the derivative of $g$ is an absolutely continuous function on $[\epsilon,\infty)$. 
  \end{lem}
\textbf{Proof:} 
The facts that $g$ is non-decreasing and $g(0)=0$ are obvious.
Clearly, for any $\lambda\in (0,1)$ and $x,y\geq 0$ we have 
$\lambda\mathcal A_{x}+(1-\lambda)\mathcal A_y\subset \mathcal A_{\lambda x+(1-\lambda)y}$
and so the concavity of $g$ follows from the concavity of the map
$\phi\rightarrow \mathbb E\left[V^{\phi,S_0}\right]$. 

Let us show that $g'(0)=1$. 
For a given $\epsilon>0$ consider the deterministic trading strategy 
$\phi\in\mathcal A_{\epsilon}$ defined by
$\phi_t:=-2 t \mathbb I_{t<\sqrt\epsilon}$ where $\mathbb I$ denotes the indicator function. 
Then, from the super-martingale property of $S$ we get 
$$g(\epsilon)\geq\mathbb E\left[V^{\phi,1}\right]\geq\epsilon \mathbb E[S_{\sqrt\epsilon}]-\frac{\Lambda}{2}\int_{0}^{\sqrt\epsilon}\phi^2_t dt= \epsilon e^{\mu\sqrt\epsilon}-\frac{2\Lambda \epsilon^{3/2}}{3}.
$$
This together with the inequality $g(\epsilon)\leq \epsilon$ (follows from (\ref{1+}))
gives $g'(0)=1$. 

Next, since $g$ is concave, the left and right derivatives $g'_{-}$ and $g'_{+}$ 
are well defined, non-increasing functions which 
satisfy $g'_{+}\leq g'_{-}$. 
Thus, in order to establish the absolute continuity (on $[\epsilon,\infty)$, $\epsilon>0$) of the derivative it is sufficient to prove 
that for any $0<x<y<2 x$ we have 
\begin{equation}\label{10}
g'_{-}(x)-g'_{+}(y)\leq \frac{4}{x} (y-x).
\end{equation}
To this end choose $0<x<y<2 x$.
Let $\hat\phi\in\mathcal A_{\frac{x+y}{2}}$ be the optimal strategy (which exists by Lemma \ref{lem1}), i.e. 
 $g\left(\frac{x+y}{2}\right)=\mathbb E\left[V^{\hat\phi,1}\right]$. 
 Define 
$\hat\phi_1\in\mathcal A_{x-\frac{y-x}{2}}$ by 
$\hat\phi_1:=\frac{3x-y}{x+y}\hat\phi$ and 
 $\hat\phi_2\in\mathcal A_{y+\frac{y-x}{2}}$
 by $\hat\phi_2:=\frac{3y-x}{x+y}\hat\phi$.
 From (\ref{1+}) and the fact that $g\geq 0$ we obtain 
\begin{equation}\label{10+}
\mathbb E\left[\frac{\Lambda}{2}\int_{0}^{\infty} \hat\phi^2_t dt\right]\leq \frac{x+y}{2}.
\end{equation}
Thus,
\begin{align*}
&\frac{y-x}{2}\left(g'_{-}(x)-g'_{+}(y)\right)\\
&\leq
\left(g\left(x\right)-g\left(x-\frac{y-x}{2}\right)\right)-\left(g\left(y+\frac{y-x}{2}\right)-g\left(y\right)\right)\\
&\leq 2 \left(g\left(\frac{x+y}{2}\right)-\frac{g\left(x-\frac{y-x}{2}\right)+g\left(y+\frac{y-x}{2}\right)}{2}\right)\\
&\leq \frac{1}{2}\mathbb E\left[2 V^{\hat\phi}-V^{\hat\phi,1}-V^{\hat\phi,2} \right] \\
&=\left(\left(\frac{3x-y}{x+y}\right)^2+\left(\frac{3y-x}{x+y}\right)^2-2\right)\mathbb E\left[\frac{\Lambda}{2}\int_{0}^{\infty} \hat\phi^2_t dt\right]\\
&\leq \frac{4 (y-x)^2}{x+y}
\end{align*}
The first two inequalities are due to the concavity of $g$. The third inequality
follows from the optimality of $\hat\phi\in\mathcal A_{\frac{x+y}{2}}$. The equality is straightforward. 
The last inequality follows from (\ref{10+}) and simple arithmetics. We conclude (\ref{10}) and complete the proof. 
\qed
${}$\\
${}$\\
We arrive at the following auxiliary result.
\begin{lem}\label{lem3}
Let $f:[0,\infty)\rightarrow [0,\infty)$ be a function which satisfies 
properties (i)-(ii) in Theorem \ref{thm2.1}. There exists a unique trading strategy $\tilde\phi\in\mathcal A_{\Phi_0}$ which solves the ODE
\begin{equation}\label{11}
\tilde\phi_t:=\frac{d\tilde\Phi_t}{dt}=-\frac{1}{\Lambda}S_t\left(1-f'\left(\frac{\tilde\Phi_t}{S_t}\right)\right), \ \ \ t\geq 0.
\end{equation}
Moreover, the process 
$M_t:=S_t+\lambda\tilde\phi_t=S_t f'\left(\frac{\tilde\Phi_t}{S_t}\right)$, $t\geq 0$
is a non-negative 
super-martingale which satisfies 
\begin{equation}\label{12}
\mathbb E\left[-\int_{0}^{\infty}\tilde\phi_t M_tdt\right]=\Phi_0 M_0.
\end{equation} 
\end{lem}
\textbf{Proof:}
From properties (i)-(ii) in Theorem \ref{thm2.1}
it follows that the function $f':(0,\infty)\rightarrow[0,1]$ is locally Lipschitz continuous, and so there exists 
a unique solution (see Chapter II in \cite{W}) to the ODE (\ref{11}) up to the hitting time $\tau:=\inf\{t:\tilde\Phi_t=0\}$. 
After this hitting time we set $\tilde\phi\equiv 0$.

Next, the fact that $f'\geq 0$ implies that the process $M$ is non-negative.
Let us show that $M_t$, $t\in [0,\tau)$ is a local-martingale.
From the integration by parts formula, (\ref{1}) and (\ref{11}) it follows that 
\begin{equation}\label{13}
d\left(\frac{\tilde\Phi_t}{S_t}\right)=-
\left(\frac{1}{\Lambda}\left(1-f'\left(\frac{\tilde\Phi_t}{S_t}\right)\right)+\left(\mu-\sigma^2\right)\frac{\tilde\Phi_t}{S_t}\right)dt-
\sigma\frac{\tilde\Phi_t}{S_t} dW_t.
\end{equation}
From (\ref{4}) it follows that $f\in C^{\infty}(0,\infty)$ and so, by applying the It\^{o} formula,
(\ref{1}) and (\ref{13}) we obtain that for $t<\tau$ the drift of $M_t$ is given by
\begin{align*}
&\mu S_t f'\left(\frac{\tilde\Phi_t}{S_t}\right)-S_t f''\left(\frac{\tilde\Phi_t}{S_t}\right)
\left(\frac{1}{\Lambda}\left(1-f'\left(\frac{\tilde\Phi_t}{S_t}\right)\right)+\left(\mu-\sigma^2\right)\frac{\tilde\Phi_t}{S_t}\right)\\
&+\frac{1}{2}\sigma^2 S_t\left(\frac{\tilde\Phi_t}{S_t}\right)^2 f'''\left(\frac{\tilde\Phi_t}{S_t}\right)-\sigma^2 
S_t \frac{\tilde\Phi_t}{S_t} f''\left(\frac{\tilde\Phi_t}{S_t}\right)\\
&=S_t\left(\frac{1}{2}\sigma^2 x^2 f'''(x)-\mu x f''(x)+\mu f'(x)-(1-f'(x))f''(x)/\Lambda\right)_{x=\frac{\tilde\Phi_t}{S_t}}.
\end{align*}
By differentiating (\ref{4}) we obtain
$$\frac{1}{2}\sigma^2 x^2 f'''(x)-\mu xf''(x)+\mu f'(x)-(1-f'(x))f''(x)/\Lambda=0$$
and so $M_t$, $t\in [0,\tau)$ is a local-martingale.
Since $M_t=S_t$ for $t\geq \tau$ we conclude that $M$ is a super-martingale.  

Finally, we establish (\ref{12}). 
Choose a sequence $a_n>0$, $n\in\mathbb N$ such that 
$$\lim_{n\rightarrow\infty}\mathbb E\left[\max_{0\leq t\leq n}S_t \left(\frac{1}{a_n}+\mathbb I_{\max_{0\leq t\leq n}S_t\geq a_n}\right)\right]=0.$$
Define a sequence of stopping times
$$\tau_n:=n\wedge\inf\{t: S_t\geq a_n\}\wedge\inf\{t:\tilde\Phi_t\leq 1/a_n\}, \ \ n\in\mathbb N.$$ 
Observe that
$$
\mathbb E\left[\tilde\Phi_{\tau_n}M_{\tau_n}\right]\leq \Phi_0\mathbb E\left[S_n\right]+\Phi_0\mathbb E\left[\max_{0\leq t\leq n}S_t \mathbb I_{\max_{0\leq t\leq n}S_t\geq a_n}\right] 
+\frac{1}{a_n}\mathbb E\left[\max_{0\leq t\leq n}S_t\right].
$$
Hence, (recall that the drift $\mu<0$) 
\begin{equation}\label{14}
\lim_{n\rightarrow\infty}\mathbb E\left[\tilde\Phi_{\tau_n}M_{\tau_n}\right]=0.
\end{equation}
We get, 
\begin{align*}
&\mathbb E\left[-\int_{0}^{\infty}\tilde\phi_t M_tdt\right]
=\mathbb E\left[-\int_{0}^{\tau}\tilde\phi_t M_tdt\right]=
\lim_{n\rightarrow\infty}
\mathbb E\left[-\int_{0}^{\tau_n}\tilde\phi_t M_tdt\right]\\
&
=\Phi_0 M_0-\lim_{n\rightarrow\infty}\mathbb E\left[\tilde\Phi_{\tau_n}M_{\tau_n}\right]=\Phi_0M_0.
\end{align*} 
The first equality is due to the fact that 
$\tilde\phi_t=0$ for $t\geq \tau$. The second equality follows from the 
monotone convergence theorem and the relation $\tau_n\uparrow\tau$ as $n\rightarrow\infty$.
From the fact that $M_{[0,\tau)}$ is a local martingale and $M\leq S$ we conclude that 
$M_{t\wedge\tau_n}$, $t \geq 0$ is a martingale. This gives the third equality. The last equality follows from (\ref{14}).
\qed
${}$\\
${}$\\
We now have all the pieces in place that we need for the
\textbf{\textit{completion of the proof of Theorem \ref{thm2.1}}}.\\
${}$\\
\textbf{Proof:}
First, we show that $g$ solves the ODE (\ref{4}). Observe that it is sufficient to prove 
that (\ref{4}) holds Lebsegue almost everywhere. Indeed, assume that (\ref{4}) holds almost everywhere. 
Choose $z>0$ for which (\ref{4}) holds. 
Then from the fact that $g'$ is continuous (due to Lemma \ref{lem2}) we obtain that
$g''(x)=\frac{d}{dx}\int_{z}^{x} g''(y) dy$ for all $x$ 
and so (\ref{4}) holds for all $x$.

Let us show that (\ref{4}) holds Lebsegue almost everywhere. For any $\phi\in\mathcal A_{\Phi_0}$
introduce the stochastic process 
$$N^{\phi}_t:=-\int_{0}^t \phi_u S_u-\frac{\Lambda}{2}\int_{0}^t \phi^2_u du+ S^2_t g\left(\frac{\Phi_t}{S_t}\right), \ \ t\geq 0.$$
From Lemma \ref{lem2} it follows that $g''$ exists 
Lebsegue almost everywhere and so we can apply the It\^{o} formula (see Problem 7.3 in \cite{KS})
and compute the drift of the processes $N^{\phi}$.
Set $\psi_t:=\frac{\phi_t}{S_t}$, $t\geq 0$.
 Similar computations as in Lemma \ref{lem3} yield that on the event $\{\Phi>0\}$ the drift of
$N^{\phi}$ is given by $(dt\otimes\mathbb P$ a.s.)
\begin{align}\label{15}
&-S^2_t\left(\psi_t+\frac{\Lambda}{2}\psi^2_t\right)+\left(2\mu-\sigma^2\right)S^2_t g\left(\frac{\Phi_t}{S_t}\right)\nonumber\\
&+S^2_t g'\left(\frac{\Phi_t}{S_t}\right)\left(\psi_t+\left(\sigma^2-\mu\right)\frac{\Phi_t}{S_t}\right)\nonumber\\
&+\frac{1}{2}\sigma^2 S^2_t \left(\frac{\Phi_t}{S_t}\right)^2 g''\left(\frac{\Phi_t}{S_t}\right)-2\sigma^2 S^2_t \frac{\Phi_t}{S_t}g'\left(\frac{\Phi_t}{S_t}\right)\nonumber\\
&=S^2_t\left(\frac{1}{2}\sigma^2x^2 g''(x)-(\mu+\sigma^2) xg'(x)+(2\mu-\sigma^2) g(x)\right)_{x=\frac{\Phi_t}{S_t}}\nonumber\\
&+S^2_t\left(\psi_t g'\left(\frac{\Phi_t}{S_t}\right)-\psi_t-\frac{\Lambda}{2}\psi^2_t\right)
\end{align} 
where the 
last equality is obvious. From the dynamic programming principle we obtain that the drift of $N^{\phi}$ is non-negative and 
is equal to zero for the case where $\phi$ is the unique optimal control (which exists by Lemma \ref{lem1}).
By looking at the quadratic form
$\psi_t\rightarrow \psi_t g'\left(\frac{\Phi_t}{S_t}\right)-\psi_t-\frac{\Lambda}{2}\psi^2_t$
we conclude that for the unique optimal control $\hat\phi\in\mathcal A_{\Phi_0}$ we have 
$$\hat\phi_t:=\frac{d\hat\Phi}{dt}=-\frac{1}{\Lambda}S_t\left(1-g'\left(\frac{\hat\Phi_t}{S_t}\right)\right), \ \ t\geq 0.$$
This together with (\ref{15}) and the fact that the drift of $N^{\hat\phi}$ is zero yields that (\ref{4}) holds true 
Lebesgue almost surely. 

Finally, we prove the uniqueness property in Theorem \ref{thm2.1}. 
Let $f:[0,\infty)\rightarrow [0,\infty)$ be a function which satisfies 
properties (i)-(ii) in Theorem \ref{thm2.1}. Consider the trading strategy $\tilde\phi\in\mathcal A_{\Phi_0}$ 
which is given by Lemma \ref{lem3}. Since the optimal trading strategy is unique, in order to prove 
the uniqueness property in Theorem \ref{thm2.1} it is sufficient to show that $\tilde\phi\in\mathcal A_{\Phi_0}$ is optimal. 
Choose an arbitrary $\phi\in\mathcal A_{\Phi_0}$. 
Recall the non-negative super-martingale $M$ from Lemma \ref{lem3}.
Then,
\begin{align*}
&\mathbb E\left[V^{\tilde\phi,S_0}\right]=\Phi_0M_0+\mathbb E\left[\int_{0}^{\infty}\tilde\phi_t (M_t-S_t) dt-\frac{\Lambda}{2}\int_{0}^{\infty} \tilde\phi^2_t dt\right]\\
&\geq \Phi_0M_0+\mathbb E\left[\int_{0}^{\infty}\phi_t (M_t-S_t) dt-\frac{\Lambda}{2}\int_{0}^{\infty} \phi^2_t dt\right]\geq\mathbb E\left[V^{\phi,S_0}\right].
\end{align*}
The quality is due to (\ref{12}).
The first inequality follows from the fact that the quadratic form 
$\phi (M-S)-\frac{\Lambda}{2}\phi^2$ takes its maximum for $\phi=\tilde\phi$. 
The last inequality follows from the fact that $M$ 
is a non-negative super-martingale. This completes the proof. 
\qed

\section*{Acknowledgments}
I would like to thank Eyal Neuman for fruitful discussions. 
This research was partially supported by the ISF grant 230/21.

\end{document}